\documentclass[prb,preprint,aps,showpacs]{revtex4}
\usepackage{graphics}
\usepackage{graphicx}
\usepackage{amssymb}
\usepackage{epsfig}
\usepackage{color}
\usepackage{ulem}
\usepackage{arydshln}
\usepackage{multirow}

\newlength{\upit}\upit=0.1truein

\newcommand{\ltappr}{{{\lower4pt\hbox{$<$} } \atop \widetilde{ \ \ \ }}}
\newlength{\bxwidth}\bxwidth=1.5 truein

\newlength{\figwidth}
\newlength{\shift}
\newcommand \bea {\begin{eqnarray} }
\newcommand \eea {\end{eqnarray}}

\begin{document}

\title{Physics of Polymorphic Transitions in CeRuSn}
\author{
J.\ Fik\'{a}\v{c}ek$^\ast$, J.Prokle\v{s}ka, M.\ M\'{i}\v{s}ek, J.
Custers, S.\ Dani\v{s}, J.\ Prchal, V.\ Sechovsk\'{y}}

\affiliation{Faculty of Mathematics and Physics, Charles
University in Prague, Ke Karlovu 5, 121 16 Prague 2, Czech
Republic}

\author{I. C\'{i}sa\v{r}ov\'{a}}
\affiliation{Department of Inorganic Chemistry, Faculty of
Science, Charles University in Prague, Hlavova 8, 128 43 Prague 2,
Czech Republic}


\begin{abstract}
We report a detailed study of the polymorphic transitions in
ternary stannide CeRuSn on high quality single crystals through a
combination of X--ray diffraction experiments conducted at 300,
275 and 120~K, and measurements of the thermal expansion,
magnetization, and resistivity, along main
crystallographic axes. In addition, the transition was followed as
a function of pressure up to 0.8~GPa. The present X--ray
diffraction data show that the room temperature polymorph consists
of the lattice doubled along the $c$ axis with respect to the
CeCoAl-type structure consistent with previous reports. Upon
cooling, the compound undergoes two successive transitions, first
to a quintuple ($\sim 290$~K) and than to a triple CeCoAl
superstructure at $\sim 225$~K. The transitions are accompanied by
a tremendous volume change due to a strong shrinking of the
lattice along the $c$ axis, which is clearly observed in thermal
expansion. We advance arguments that the volume collapse
originates from an increasing number of crystallographically
inequivalent Ce sites and the change of ratio between the short
and long Ce--Ru bonds. The observed properties of the polymorphic
transition in CeRuSn are reminiscent of the $\gamma \rightarrow
\alpha$ transition in elementary Cerium, suggesting that similar
physics, i.e.\,, a Kondo influenced transition and strong lattice
vibrations might be the driving forces.

\end{abstract}
\date{\today}

\maketitle
\newcommand{\CRS}{CeRuSn}
\newcommand{\LIS}{LaIr$_2$Si$_2$}
\newcommand{\Tcl}{$T^{\mathrm{c}}_{\mathrm{L}}$}
\newcommand{\Tcu}{$T^{\mathrm{c}}_{\mathrm{U}}$}
\newcommand{\Twl}{$T^{\mathrm{w}}_{\mathrm{L}}$}
\newcommand{\Twu}{$T^{\mathrm{w}}_{\mathrm{U}}$}

\section{Introduction}
The 4{\it f\/} electron in Cerium is energetically weakly bound
despite the fact that it resides deep within the core of the atom.
This is due to the relatively extended nature of the 4{\it f\/}
wave function. Strong correlations between the Ce 4{\it f\/}
electron and hybridization between the 4{\it f\/} state with
those of the ligand states arise in Ce compounds and hence the
local environment of the Ce-atom dictates whether this electron
will remain bound within the core (Ce$^{3+}$ state), join the
spatially extended valence electrons (Ce$^{4+}$ state), or reside
with certain probability in each. In the last case, the atom is
said to be in an ``intermediate valent" state with fluctuating
charge occupancy of the 4{\it f\/} shell
state.\cite{Sereni1982,Johansson1987,Malterre1989} The fragile
character of the 4{\it f\/} electron due to its sensitivity on
interatomic distances, as this determines the hybridization
strength, is capitalized on for example investigating quantum
critical phenomena, where by either applying hydrostatic pressure
or chemical substitution the unit-cell volume shrinks or expands
with the result that ground state of the material under
investigation changes.\cite{Loehn2007}\\
Knowledge about the electronic structure and understanding its
relation to the physical properties observed in intermetallics, and
in particular rare earth based compounds is an ambitious
undertaking in condensed matter research. Investigation by varying
the properties utilizing pressure or chemical substitution is one
way. In addition, some compounds exist in more than one
crystallographic structure.\cite{Ghad1988} By means of pressure
and/or temperature it is possible to convey one into the other
reversibly. Such polymorphic transitions allow for comparative
study.\cite{Mihalik2010} The chemical composition is retained but
bonds of the individual atoms and therefore, the overall
electronic structure and related physical properties can differ
significantly. An example is \LIS . Polymorphism of \LIS\ between
a high-temperature phase of the primitive tetragonal
CaBe$_2$Ge$_2$-type structure and a low-temperature phase of the
body-centered tetragonal ThCr$_2$Si$_2$-type structure has been
demonstrated by Braun {\it et al.}.\cite{Braun1983} Notably, the
high-temperature modification displays superconductivity below
1.6~K, while the low-temperature phase is
normal down to 1~K.\\

The polymorphic isostructural $\gamma$ (fcc) $\rightarrow \alpha$
(fcc) transition in elementary Cerium~\cite{Bridgman1948} is
illustrative for the difficulty in understanding the complexity
between ``chemical bonds", ``physical properties" and ``structural
transition" especially in materials with electrons near the
boundary between itinerant and localized
behavior.\cite{Koskenmake1978} The transition involves a large
volume collapse of $\sim 17$~\% at room temperature and pressure
$\sim 0.8$~GPa. A general consensus exists to attribute the
transition to an instability of the Ce 4{\it f\/} electron.
However, Johansson~\cite{Johansson1974} explained the $\gamma
\rightarrow \alpha$ transition as sort of Mott transition in which
the localized 4{\it f\/} electrons in the $\gamma$-phase become
itinerant and participate in bonding in the lower volume $\alpha$-phase.
This model continues to compete with the
Kondo-volume-collapse scenario,\cite{Allen1982,Allen1992} which
assumes that the 4{\it f\/} electron is localized in both the
$\gamma$- and $\alpha$-phases. The loss of magnetic moment in the
$\alpha$-phase results from screening of the moments by the
surrounding conduction electrons. To complicate, latest neutron
and X--ray diffraction studies acknowledge the importance
of lattice vibrations as well.\cite{Jeong2004,Lipp2008}\\

In many aspects the recently observed polymorphic transition in
the equiatomic stannide \CRS\ seems to have much in common with
the $\gamma \rightarrow \alpha$
transition in Cerium. \\
\CRS\ at room temperature crystallizes in a superstructure
modification of the monoclinic CeCoAl-type crystal structure (new
monoclinic type, space group C2/$m$) with lattice parameters $a=
11.561(4)$~\AA\,, $b= 4.759(2)$~\AA\,, $c= 10.233(4)$~\AA\,, and
$\beta = 102.89(3)^\circ$.\cite{Riecken2007} As a consequence of
this doubling of the original CeCoAl unit cell along the $c$ axis,
the compound possesses two crystallographic independent cerium
sites labelled Ce1 and Ce2. Although topology of both Cerium sites
is identical, five rhodium, six tin, and six cerium atoms in the
coordination shell, the tiny changes in interatomic distances,
most notably the Ce--Ru bonds (Ce1--Ru: ranging from 2.33 to
2.46\AA\,; Ce2--Ru: ranging from 2.88 to 2.91~\AA\,) result in Ce1
being in intermediate valent state while Ce2 shows strong
localization of the {\it f\/} electron as suggested by magnetic
susceptibility experiments.\cite{Riecken2007,Mydosh2011} This
presumption is borne out by electronic structure
calculations~\cite{Matar2007} and proven by X--ray absorption
near-edge structures (XANES) data.\cite{Feyerherm2012}
Latest yields average valencies of 3.18 for Ce1 and Ce2. \\
The polymorphic transition in \CRS\ sets in just below room
temperature at $\sim 290$~K and is completed at around $\sim
160$~K upon cooling. The reverse transformation occurs on heating
with $\sim 170$~K and $\sim 320$~K as the onset and end
temperatures, respectively. Initial measurements of the magnetic
susceptibility, specific heat, thermopower, and resistivity were
performed on polycrystalline samples.\cite{Mydosh2011} The
transformation was smeared out and manifested as broad hysteresis
with a cusp-like structure in resistivity, a step-like decrease of
the susceptibility, a broad hump in the specific heat and a strong
increase in thermopower. A detailed analysis of the transition by
means of synchotron X--ray diffraction experiments on a single
crystal revealed that the room temperature phase is replaced by a
set of close to commensurate modulations along the $c$ axis,
namely quintupling ($\sim 290$~K) and (dominant) quadrupling
(below 210~K) before finalizing ($\sim 180$~K) in an ill-defined
modulated ground state, which is close to a tripling of the basic
monoclinic CeCoAl-type structure.\cite{Feyerherm2012} \\

The present work gives a detailed examination of the physical
properties of the polymorphic transition of \CRS\. For this
purpose, measurements were performed on high quality single
crystals. The lower amount of crystal lattice defects, absence of
grain boundaries and the ability to perform experiments along
specific crystallographic orientations allows us to resolve
details related to the transition and to attribute those
signatures in the experiments to the respective modulation in the
structure.

\section{Experimental Details}
\label{Sec1}
\subsubsection*{Sample preparation}
Single crystals of \CRS\ were prepared in two stages. First a
polycrystalline button of the nominal 1:1:1 stoichiometry was
synthesized using elements of purity 3N Ce (Ce from Alpha Aesar
which was additionally purified by solid state electrotransport
technique\cite{Carlson1977}), 4N Ru and 5N Sn as starting
materials. The reaction of the stoichiometric mixture of the
elements was performed on a water-cooled copper crucible in a
mono-arc furnace under 6N Argon atmosphere. The mass difference
before and after the reactions was negligible ($< 0.1$~\%). The
crystal was than grown utilizing a modified Czochralski technique;
the button was remelted in a tri--arc furnace under 6N Argon
protection atmosphere and a tungsten rod was used as a seed.\\
The quality of the single crystal was checked by X--ray Laue
back--scattering, which was also used for orienting the crystals
later on. The chemical composition was verified employing a Tescan
Mira I LMH scanning electron microscope (SEM). The instrument is
equipped with a Bruker AXS energy dispersive X--ray detector
(EDX). Within the accuracy of the device, no impurity phases were
resolved and the measurement confirmed the correct
1:1:1 stoichiometry. \\
Afterwards, the crystal was cut for further analysis. One piece
was pulverized and examined at room temperature by means of powder
X--ray diffraction (Bruker D8 Advance diffractometer with
Cu-K$_\alpha$ radiation with $\lambda = 1.5405$~\AA\,). The
obtained diffraction patterns were refined by Rietveld analysis
using FULLPROF.\cite{Fullprof} The analysis confirmed the CeCoAl
superstructure and the corresponding lattice parameters agreed
well with those values reported in
literature.\cite{Riecken2007}\\
The other piece of the crystal was annealed at 700~$^\circ$C for
one week in vacuum ($p= 1 \times 10^{-6}$~mbar) in order to
improve homogeneity. In the following, the whole characterization
procedure was repeated unveiling no significant differences.\\

\subsubsection*{Experimental setup}
From the annealed single crystal a small piece was cut for
investigating the crystal structure by X--rays at defined
temperatures. Therefore, the approximately $0.1 \times 0.1 \times
0.1$~mm$^3$ piece was placed inside a Lindemann capillary. The
capillary itself was mounted into a Bruker Apex II diffractometer
with Mo--K$_\alpha$ radiation ($\lambda = 0.71073$~\AA\,). In
order to reach lower temperatures, the capillary was inserted into
a flow of cold nitrogen gas. The crystal structure was resolved by
direct methods~\cite{Sheldrick2007} and adjacent refinement was
done by
full--matrix least--squares based on $F^2$. \\

Bulk properties were retrieved employing standard equipment. The
magnetization was measured in a MPMS7 (Quantum Design). Data were
collected in the temperature range from 1.8 to 350~K and in fields
up to 7~T. Resistivity, Hall resistivity, thermopower, thermal
conductivity were measured in a PPMS14 (Quantum Design) using the
respective optional accessories of the device. The temperature was
varied between 1.8 and 350~K and magnetic fields up to 14~T were
applied. The resistivity was measured using standard 4--point
technique. In order to reduce contact resistance, the 25~$\mu$m
diameter Au--wires were spot welded onto the sample. Measurements
of the resistivity were performed at ambient and hydrostatic
pressure. For the later, the PPMS device was used only to control
temperature. The sample was loaded into a double cylinder
CuBe/NiCrAl pressure cell. Daphne 7373 oil was used as pressure
medium and the applied pressure was determined at
room temperature utilizing a manganin manometer. \\
The thermal expansion was measured in a temperature interval of
180--340~K. The sample was built into a miniature capacitance
cell.\cite{Rotter1998} The capacity was read out by an Andeen
Hagerling 2500A capacitance bridge. The cell was inserted into the
PPMS whose controlling was used to set temperature. \\

Most of the experiments were conducted on both, the as cast and
the annealed single crystals. The quality of the crystals improved
considerably by annealing. The resistivity
behavior of the as cast crystals to some extent resembled the
results of the polycrystalline sample presented in earlier
work\cite{Mydosh2011} that is a single broad hysteresis, which
differs in detail depending on the current
direction with respect to the crystallographic axis. On the
contrary the annealed crystals exhibit two sharp transitions
evident for two distinct transitions as will be discussed below.
In addition, differences in the hysteresis of each of the
transitions could be resolved. Results presented in this report
were obtained on the annealed crystals. The bulk properties were
measured with respect to the three principle crystallographic
axes. Data shown have been collected on two batches. The
resistivity and susceptibility experiments were performed on
batch I, while for thermal expansion a piece from batch II was used.
Hence, slight differences in the respective transition
temperatures are observed, which we attribute to sample
dependencies.

\section{Results}
\subsection{Measurement of bulk properties}
\begin{figure}[t]
\centerline{\includegraphics[width=0.8\textwidth]{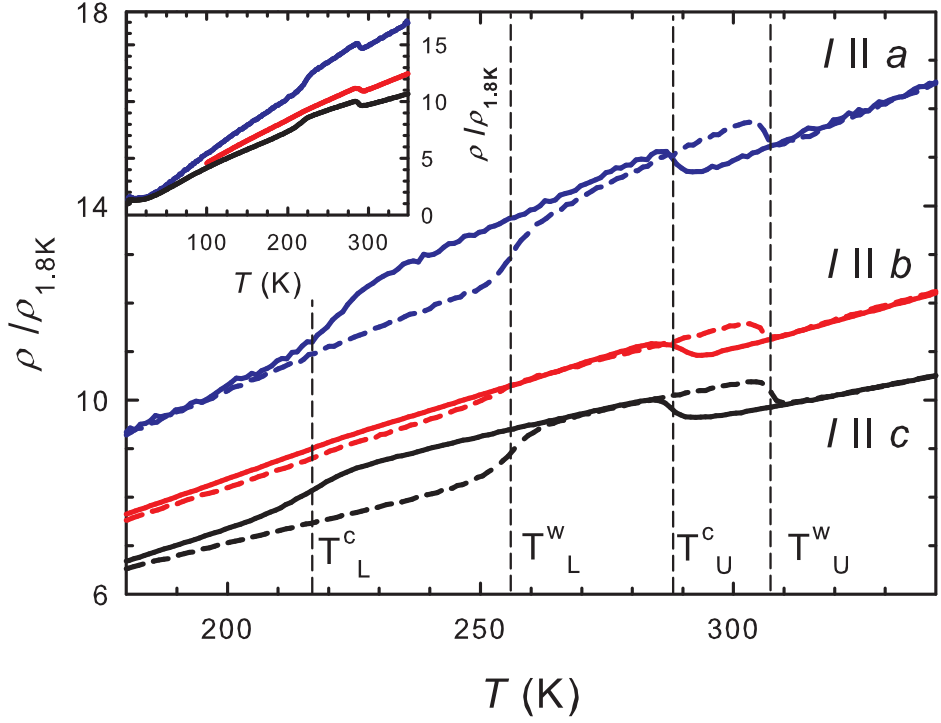}}
%\centerline{\includegraphics[width=\columnwidth]{FikacekCeRuSnFigure1.eps}}
\caption{\label{Fig1}Temperature dependences of the relative
resistivity $\rho(T)/\rho_{1.8\mathrm{K}}$ of CeRuSn measured with
current applied along the 3 principle crystallographic directions
for cooling (solid lines) and warming (short--dashed lines). The
main panel focuses on the temperature range the discussed
transitions take place. The vertical dashed lines mark the
estimated transition temperatures \Tcu\,, \Tcl\,, \Twl\, and
\Twu\,, respectively (see text). The inset shows the whole
temperature range of the experiment for cooling regime.}
\end{figure}

Figure~\ref{Fig1} depicts the temperature dependence of the scaled
electrical resistivity $\rho/\rho_{1.8\mathrm{K}}$ for current
applied along the $a$, $b$ and $c$ axis, respectively. The inset
shows the full temperature range. The different corresponding
values for $\rho(T)$ document anisotropy of the electronic
transport in \CRS\,. Yielding
$\rho_{300\mathrm{K}}/\rho_{1.8\mathrm{K}} \sim 18$ (for $I
\parallel a$) comparing to only $\sim 1.75$ for a polycrystalline sample proofs the high quality of our single
crystals. More remarkably, the observed behavior in resistivity
differs significantly from previously published work on a
polycrystalline sample.\cite{Mydosh2011} Cooling down the sample
(solid lines in Fig.~\ref{Fig1}) from above room temperature, the
resistivity undergoes a sharp step-like increase by about 7~\%
just below 290~K. The anomaly, indicated by \Tcu\ in the main
panel, is clearly seen in all applied current directions while
absent in the polycrystalline sample. Below 225~K a second
transition emerges marked by \Tcl\,, and the resistivity seems to
fall back onto the original curve from before the first transition
($I \parallel c$ and $a$). The transition is weakly pronounced for
$I \parallel b$ while decrease of the resistivity is only a
fraction of the increase at \Tcu\,. Interestingly, this \Tcl\
transition, although slightly shifted towards lower temperatures,
is observed in the polycrystal as well. However, contrary to our
data, resistivity increases. Below 3~K, resistivity reveals a
third anomaly, which can be attributed to the onset of
antiferromagnetic ordering reported earlier.\cite{Mydosh2011} In
the following, discussion on the antiferromagntic order is omitted
and focus is entirely on the
relevant temperature range of \Tcu\ and \Tcl\,. \\
Upon warming up (dashed lines in Fig.~\ref{Fig1}), both the lower,
\Twl\ $\sim 256$~K, and upper, \Twu\ $\sim 307$~K, transitions of
$\rho(T)$ preserve shape and size of the step. However, they are
observed at much higher temperatures than their corresponding
anomalies when cooling down, i.\,e.\,, exhibiting large
temperature hysteresis. In comparison, hysteresis of the
lower-temperature transition yields \Tcl\ $-$ \Twl\ $\sim 40$~K
almost double the hysteresis of the upper-temperature transition
\Tcu\ $-$ \Twu\ $\sim 20$~K. These remarkable features in \CRS\ remain intact even in magnetic fields up to 14~T. \\

%=====magnetization=====

\begin{figure}[t]
\centerline{\includegraphics[width=0.8\textwidth]{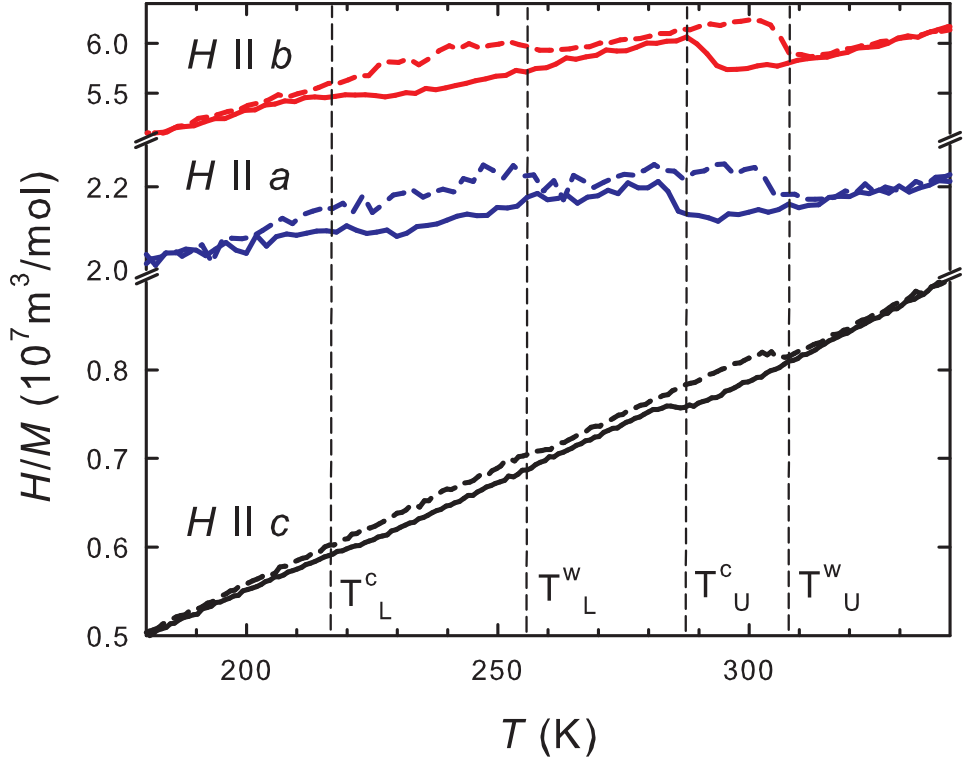}}
%\centerline{\includegraphics[width=\columnwidth]{FikacekCeRuSnFigure2.eps}}
\caption{\label{Fig2} The inverse magnetic susceptibility
$\chi^{-1}(T)= H/M(T)$ of CeRuSn in magnetic field applied along
the $a$, $b$ and $c$ axis in the temperature interval of interest.
Solid (short--dashed) lines denote measurements in cooling down
(warming up). The applied magnetic field was 1~T for $H\parallel
a$ and $H \parallel c$ and 7~T for $H \parallel b$, respectively.
The vertical lines are guides to the eye illustrating the
estimated transition temperatures \Tcu\,, \Tcl\,, \Twl\, and \Twu\
from temperatures of the resistivity anomalies. Mind the breaks in
the scale on the vertical axis.}
\end{figure}

Figure~\ref{Fig2} plots the temperature dependence (solid lines
refer to cooling down, dashed lines for warming up sequence,
respectively) of the inverse dc magnetic susceptibility,
$\chi^{-1}$, in fields applied along the principle crystallographic
axes. The magnetic susceptibility is strongly anisotropic
apparently due to the influence of the very low--symmetry crystal
electric field (CEF) on the orbitals of the Ce--ion. The
transitions at \Tcu\ and \Tcl\ when cooling down, and \Twl\ and
\Twu\ when warming up \CRS\ are clearly witnessed by a small
negative step in the magnetization, i.\,e., a positive jump in
$\chi^{-1}$. Note that the polycrystalline sample shows only a
single step upon cooling at $T \sim
180$~K.\cite{Riecken2007,Mydosh2011} The lower transitions, \Tcl\
and \Twl\ are much weaker than the upper ones.\\
Short temperature intervals that are above \Tcu\ and between \Tcu\
and \Tcl\, prevent a meaningful qualitative analysis of the
temperature dependence of each of the separate paramagnetic
phases. Quantitatively, assuming the effective moment remains
conserved across the transitions, a reasonable assumption
recalling that XANES unveil no chance in Ce
valency,\cite{Riecken2007} the change in $\chi(T)$ implies a
shift of the paramagnetic Curie temperatures towards larger
negative values for each field direction objecting statements on
the polycrystalline sample.\cite{Riecken2007}\\

%=====thermal expansion=====
\begin{figure}[t]
\centerline{\includegraphics[width=\textwidth]{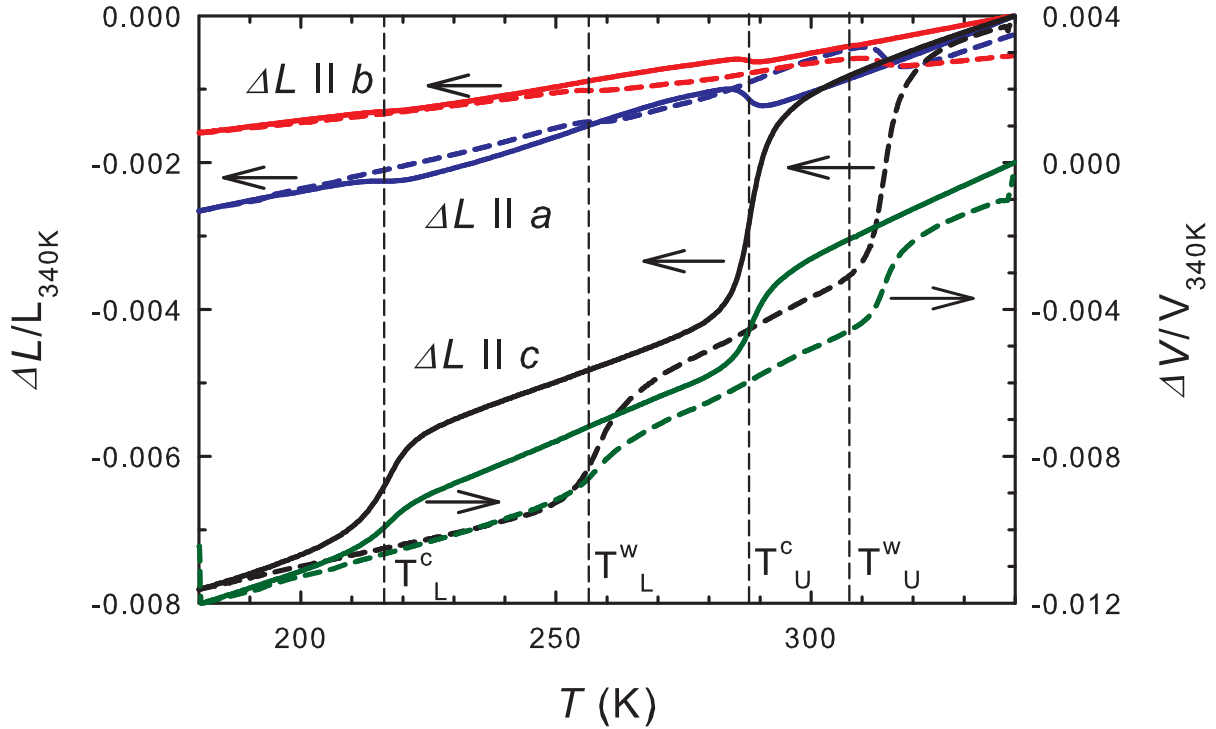}}
%\centerline{\includegraphics[width=\columnwidth]{FikacekCeRuSnFigure3.eps}}
\caption{\label{Fig3} The temperature evolution of the relative
change of length $\Delta L/L_{340 \mathrm{K}}$ (left axis)
resolved for each of the principle crystallographic directions
when the sample was in cool down (solid line) and warm up run
(short-dashed line) through the transitions. On the right axis,
the calculated relative volume change  $V/V_{340 \mathrm{K}}$ for
cooling (solid line) and warming (short--dashed line) is depicted.
% The vertical lines are
%guides to the eye illustrating the estimated transition
%temperatures TcL, TcU, TwL, and TwU, respectively, from
%temperatures of the resistivity anomalies
}
\end{figure}

Detailed dilatometric measurements performed on a well-defined
single crystal provide important information on the evolution of
the lattice parameters. The thermal expansion was measured along
each of the three principle crystallographic directions. As
presented in Fig.~\ref{Fig3}, two steps are observed along each of
the axes at similar temperatures to the anomalies recorded in
$\rho(T)$ and $\chi(T)$, respectively. When cooling (solid lines
in Fig.~\ref{Fig3}) from room temperature, the crystal contracts
considerably along the $c$ axis by almost 0.8~\% between 340 and
180~K. Contrary, the tiny positive jumps disclosed for $a$ and $b$
directions represent very small expansion of $a$ and nearly
negligible increase of the lattice parameter $b$, respectively.
Consequently, the volume changes at \Tcu\ and \Tcl\ express
mainly the $c$ axis behavior, i.\,e.\,, the crystal shrinks in two
steps with decreasing temperature. The corresponding reverse
transitions appear at \Twl\ and \Twu\ corroborating the hysteretic
behavior of the phases as inferred by resistivity and
magnetization experiments already.

%=====comparison=======
\begin{figure}[t]
\centerline{\includegraphics[width=0.8\textwidth]{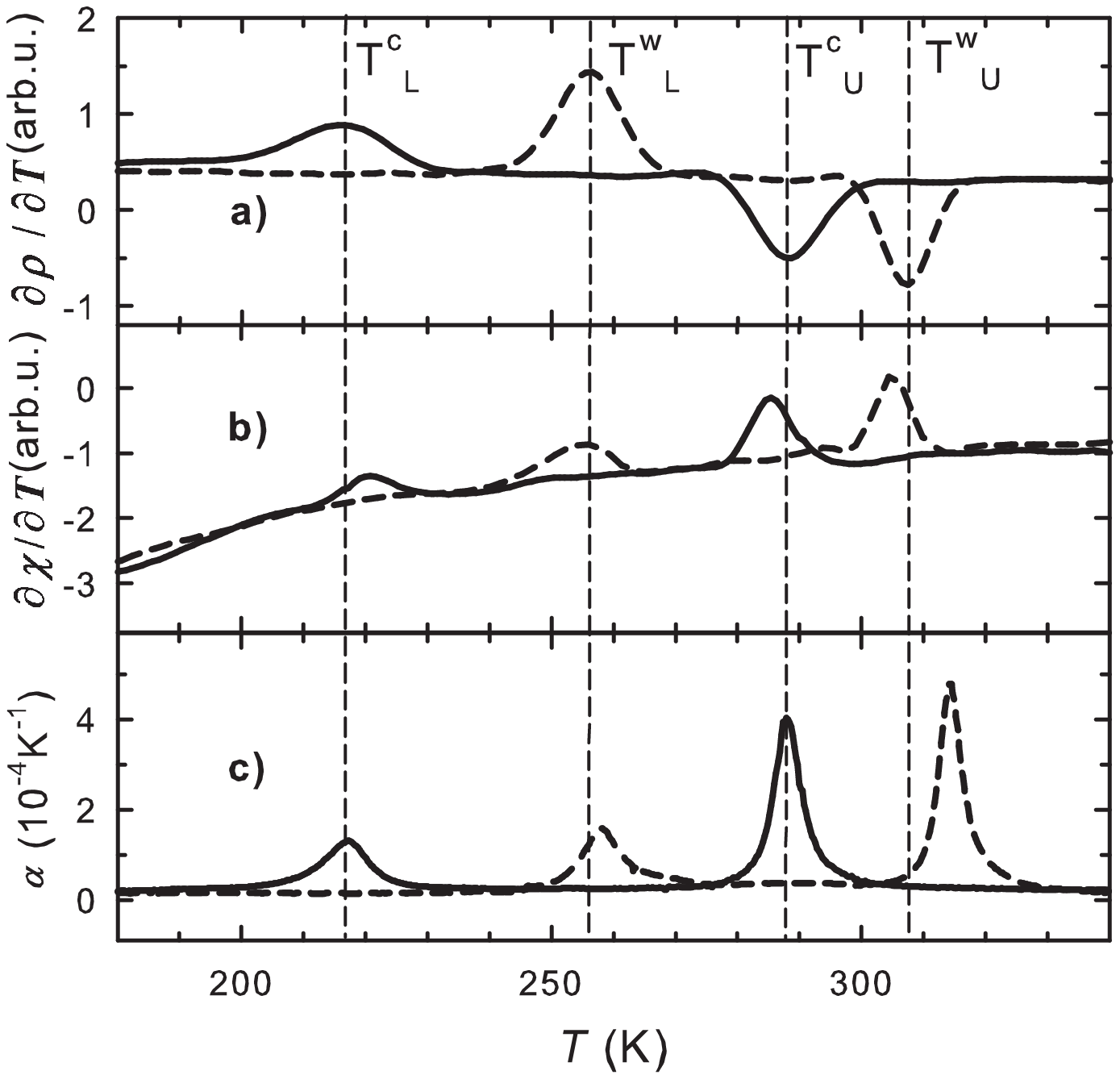}}
%\centerline{\includegraphics[width=\columnwidth]{FikacekCeRuSnFigure4.EPS}}
\caption{\label{Fig4} Comparative plot of the temperature
dependences of (a) the temperature derivative of the electrical
resistivity $\partial\rho/\partial T$ for $I \parallel c$, (b) the
temperature derivative of the magnetic susceptibility $\partial
\chi/\partial T$ with $H \parallel c$, and (c) the linear thermal
expansion coefficient $\alpha(T)$ along the $c$ axis. Solid
(short--dashed) lines show data taken when cooling down (warming
up) the sample. The vertical lines are guides to the eye
illustrating the estimated transition temperatures \Tcu\,, \Tcl\,,
\Twl\, and \Twu\,, respectively, from temperatures of the
resistivity anomalies.}
\end{figure}

In figure~\ref{Fig4}, a comparison of the temperatures of the
anomalies disclosed in aforementioned bulk experiments is made. To
visualize the location of the transition more clearly, the
temperature derivative of the resistivity ($\partial \rho/\partial
T$), dc magnetic susceptibility ($\partial \chi/\partial T$) and
thermal expansion coefficient ($\alpha$) are displayed. The maxima
in ($\partial \rho/\partial T$) and ($\alpha$) well coincide
except for the \Twu\ transition. This difference likely arose because a sample from batch II had been used,
as mentioned in
section~\ref{Sec1}.

\subsection{X--ray single crystal study of polymorphs}
\begin{figure*}[t]
\centerline{\includegraphics[width=0.8\textwidth,angle=-90]{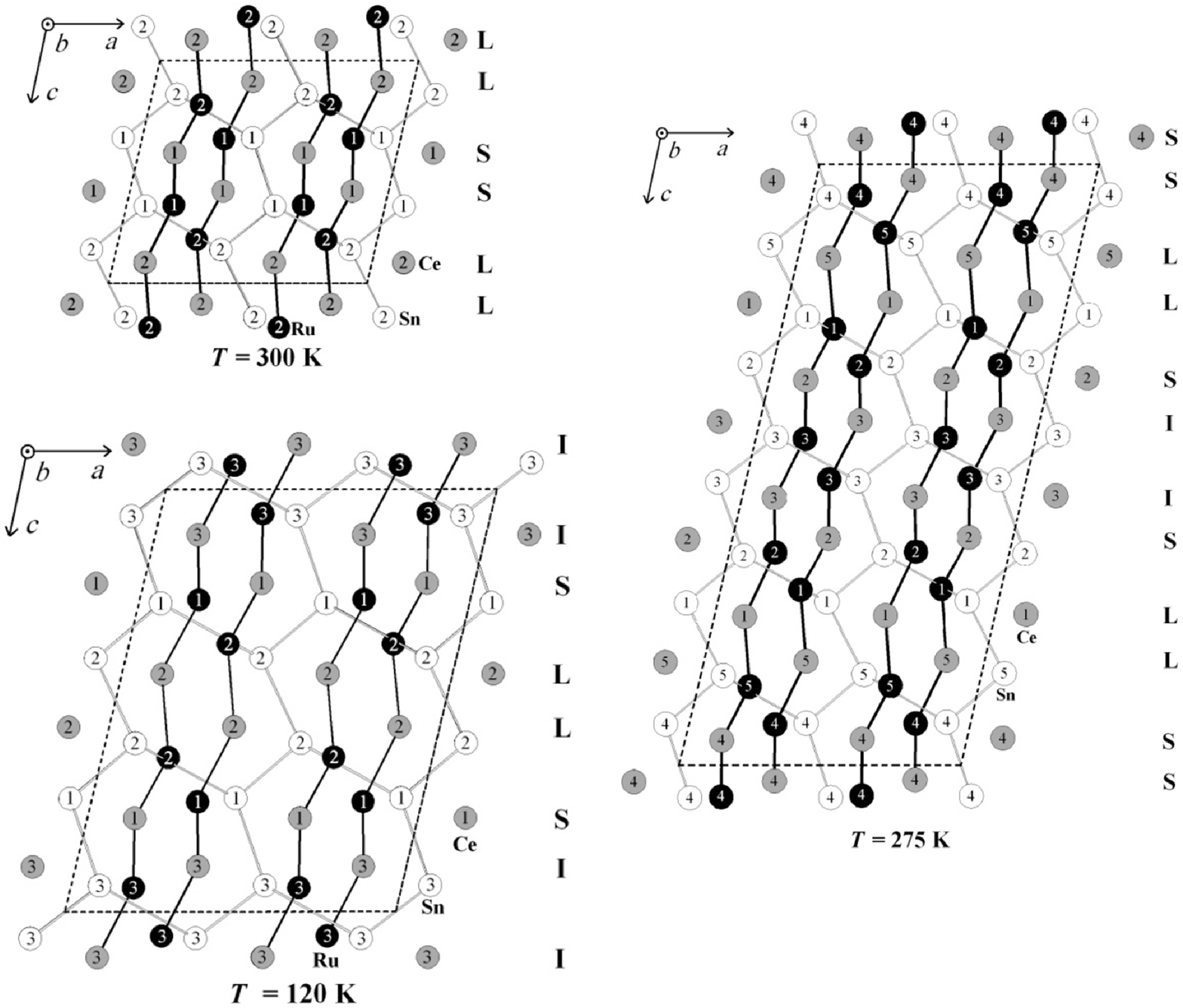}}
%\centerline{\includegraphics[width=0.7\columnwidth,angle=-180]{FikacekCeRuSnFigure5.eps}}
\caption{\label{Fig5} Illustration of CeRuSn crystal structures
above, between and below the transitions at 300, 275 and 120~K,
respectively. For all types of atoms, numbers label two
non-equivalent crystallographic sites. Ce sites are additionally
marked on the right-hand site by $S$, $L$ and $I$ letters
according to the similarity of their environment shown in table
II. Black and white lines emphasize the typical Ce--Ru shortest
interatomic distances and tin network, respectively. The dashed
line illustrates Bravais unit cell.}
\end{figure*}

\begin{table*}
    \caption{{Lattice parameters determined from SC X--ray diffraction. The lattice
    parameters are sorted according to the temperature evolution during our experiment.
    The last column shows the ratio between $\Delta c_{\mathrm{red}}$ standing for the $c$ axis modified to room
    temperature size and the room temperature $c$ axis value.
    It estimates the real change of crystal dimension along the $c$
    direction.}}
     \vspace*{2mm}
    \label{Table1}
    \renewcommand{\arraystretch}{1.2}
  \begin{tabular}{ccccccccccccc}
  \hline \hline
    $T$~(K)&$\quad$ &$a$~(\AA) & $\quad$& $b$~(\AA)&$\quad$ &$c$~(\AA) & $\quad$&$\beta$~($\deg$) & $\quad$&$V$~(\AA$^3$) &$\quad$ & $\Delta
    c_{\mathrm{red}}/c_{300\mathrm{K}}$ (\%)\\ [0.5ex]
   \hline
   297 & & 11.565(1) & & 4.7529(5) & & 10.2299(9) & & 103.028(2) & & 547.85(9) & & -- \\
   290 & & 11.560(3) & & 4.751(1) & & 10.227(3) & & 103.081(7) & & 547.1(2) & & 0.028(1) \\
   275 & & 11.576(2) & & 4.7556(7) & & 25.454(3) & & 102.959(4) & & 1365.6(3) & & 0.47(1) \\
   120 & & 11.566(2) & & 4.7477(6) & & 15.229(2) & & 103.554(4) & & 813.0(2) & & 0.70(2)
   \\ [1ex]
   200 & & 11.569(2) & & 4.7505(6) & & 15.237(2) & & 103.496(4) & & 814.3(2) & & 0.76(2) \\
   \hline \hline
  \end{tabular}
\end{table*}

\begin{table*}
    \caption{{Closest ruthenium neighbors of cerium atoms at different temperatures. ($S$), ($L$) and ($I$)
    denote bond type {\it short}, {\it long} and {\it intermediate} respectively (see also text).}}
     \vspace*{2mm}
    \label{Table2}
    \renewcommand{\arraystretch}{1.2}
  \begin{tabular}{llllllllllllllllllll}
  \hline \hline
  \multicolumn{6}{c}{$T=300$~K} & $\qquad$ & \multicolumn{6}{c}{$T=275$~K} & $\qquad$ & \multicolumn{6}{c}{$T=120$~K} \\ [0.5ex]
  \hline
  Ce1 & $\quad$ & 2.3277(9) & $\quad$ & Ru1 &                          & & Ce1 & $\quad$ & 2.8882(19) & $\quad$ & Ru5 &                         & & Ce1 & $\quad$ & 2.4325(18) & $\quad$ & Ru2 &  \\ [-1ex]
      & $\quad$ & 2.4661(11) & $\quad$ & Ru2 & \raisebox{1.5ex}{($S$)} & &     & $\quad$ & 2.9247(19) & $\quad$ & Ru2 & \raisebox{1.5ex}{($L$)} & &     & $\quad$ & 2.4345(18) & $\quad$ & Ru3 & \raisebox{1.5ex}{($S$)}
\\ [0.5ex]
  Ce2 & $\quad$ & 2.8783(10) & $\quad$ & Ru1 &  & &  Ce2  & $\quad$ & 2.4331(18) & $\quad$ & Ru1 &  & & Ce2
& $\quad$ & 2.9153(17) & $\quad$ & Ru1 &  \\ [-1ex]
      & $\quad$ & 2.9081(10) & $\quad$ & Ru2 & \raisebox{1.5ex}{($L$)} & &    & $\quad$ & 2.4380(19) & $\quad$ & Ru3 & \raisebox{1.5ex}{($S$)} & &
& $\quad$ & 2.9257(18) & $\quad$ & Ru2 & \raisebox{1.5ex}{($L$)}
\\ [0.5ex]
      & $\quad$ &  & $\quad$ &  &  & &  Ce3  & $\quad$ & 2.271(2) & $\quad$ & Ru2 & & &  Ce3
& $\quad$ & 2.2676(17) & $\quad$ & Ru1 &  \\ [-1ex]
      & $\quad$ &  & $\quad$ &  &  & &    & $\quad$ & 2.7575(19) & $\quad$ & Ru3 & \raisebox{1.5ex}{($I$)} & &
& $\quad$ & 2.7482(17) & $\quad$ & Ru3 & \raisebox{1.5ex}{($I$)}
\\ [0.5ex]
      & $\quad$ &  & $\quad$ &  &  & &  Ce4  & $\quad$ & 2.3219(19) & $\quad$ & Ru4 &  & &
& $\quad$ &  & $\quad$ &  &  \\ [-1ex]
      & $\quad$ &  & $\quad$ &  &  & &    & $\quad$ & 2.4779(18) & $\quad$ & Ru5 & \raisebox{1.5ex}{($S$)} & &
& $\quad$ &  & $\quad$ &  & \\ [0.5ex]
      & $\quad$ &  & $\quad$ &  &  & &  Ce5  & $\quad$ & 2.8800(18) & $\quad$ & Ru4 & & &
& $\quad$ &  & $\quad$ &  &  \\ [-1ex]
      & $\quad$ &  & $\quad$ &  &  & &    & $\quad$ & 2.9453(19) & $\quad$ & Ru1 & \raisebox{1.5ex}{($L$)} & &
& $\quad$ &  & $\quad$ &  & \\ [0.5ex]
  \hline \hline
  \end{tabular}
\end{table*}

\begin{table*}
    \caption{{Fraction atomic coordinates \mbox{$^x$/$_a$}, \mbox{$^y$/$_b$} and \mbox{$^z$/$_c$} at 300~K and 120~K.}}
     \vspace*{2mm}
    \label{Table3}
    \renewcommand{\arraystretch}{1.2}
  \begin{tabular}{llllllllllllllll}
  \hline \hline
   \multicolumn{7}{c}{ $T=300$~K} & $\qquad$ & $\qquad$ & \multicolumn{7}{c}{ $T=120$~K}  \\ [0.5ex]
  \hline
  Ce1 & $\quad$ & 0.1396 & $\quad$ & 0 & $\quad$ & 0.4146 & $\qquad$& $\qquad$ & Ce1 & $\quad$ & 0.6306 & $\quad$ & 0 & $\quad$ & 0.1075 \\
  Ce2 & $\quad$ & 0.1225 & $\quad$ & 0 & $\quad$ & 0.9063 & $\qquad$& $\qquad$ & Ce2 & $\quad$ & 0.3567 & $\quad$ & 0 & $\quad$ & 0.2216 \\
      & $\quad$ &        & $\quad$ &   & $\quad$ &        & $\qquad$& $\qquad$ & Ce3 & $\quad$ & 0.6226 & $\quad$ & 0 & $\quad$ & 0.4375
      \\[0.5ex]
  Sn1 & $\quad$ & 0.4265 & $\quad$ & 0 & $\quad$ & 0.3469 & $\qquad$& $\qquad$ & Sn1 & $\quad$ & 0.0858 & $\quad$ & 0 & $\quad$ & 0.9376 \\
  Sn2 & $\quad$ & 0.4043 & $\quad$ & 0 & $\quad$ & 0.8486 & $\qquad$& $\qquad$ & Sn2 & $\quad$ & 0.0906 & $\quad$ & 0 & $\quad$ & 0.6008 \\
      & $\quad$ &        & $\quad$ &   & $\quad$ &        & $\qquad$& $\qquad$ & Sn3 & $\quad$ & 0.0659 & $\quad$ & 0 & $\quad$ & 0.2688
      \\ [0.5ex]
  Ru1 & $\quad$ & 0.1827 & $\quad$ & 0 & $\quad$ & 0.6481 & $\qquad$& $\qquad$ & Ru1 & $\quad$ & 0.6784 & $\quad$ & 0 & $\quad$ & 0.2605 \\
  Ru2 & $\quad$ & 0.1983 & $\quad$ & 0 & $\quad$ & 0.1973 & $\qquad$& $\qquad$ & Ru2 & $\quad$ & 0.2992 & $\quad$ & 0 & $\quad$ & 0.3654 \\
      & $\quad$ &        & $\quad$ &   & $\quad$ &        & $\qquad$& $\qquad$ & Ru3 & $\quad$ & 0.3102 & $\quad$ & 0 & $\quad$ & 0.0572 \\
  \hline \hline
  \end{tabular}
\end{table*}

The presented results on resistivity, magnetization and thermal
expansion conflict in many ways with earlier
studies.\cite{Riecken2007,Mydosh2011} Moreover, it was mentioned
in the introduction that in \CRS\ several polymorphic transitions
gradually emerged on cooling, which gave rise to additional
reflections in the diffraction patterns obtained by synchotron
experiment.\cite{Feyerherm2012} Those superstructure reflections
can be described by nearly inverse--integer folded propagation
vectors having non-zero $c$ components only. The first transition
takes place just below room temperature, changing from a
\mbox{$^1$/$_2$}-- to a \mbox{$^1$/$_5$}--like modulation. Upon
cooling, these become partially suppressed and replaced by
\mbox{$^1$/$_4$}--like ones, which are dominant at 210~K. Finally,
a \mbox{$^1$/$_3$}--kind modulation develops having the most
intensive reflections below 180~K. This one coexists with the
aforementioned modulations down to at least
100~K.\cite{Feyerherm2012} While \Tcu\ can be attributed to the
first polymorphic transition (\mbox{$^1$/$_2$} $\rightarrow$
\mbox{$^1$/$_5$}) no evidence is found in the data for the
structural change \mbox{$^1$/$_5$} $\rightarrow$ \mbox{$^1$/$_4$},
which at least is accepted to manifest in thermal expansion being
an extremely sensitive experiment on lattice changes. To
anticipate speculations about the structure, a detailed X--ray
single crystal diffraction study over the majority of the
reciprocal lattice and at defined temperatures was conducted. The
following temperature sequence was applied: 300~K, 290~K, 275~K,
120~K and 200~K matching the regions for determining the structure
above \Tcu\, between \Tcu\ and \Tcl\, below \Tcl\ and below
\Twl\, respectively.\\
In all cases, the space group $C2/m$ for the unit cell has been
observed. However, the size of the unit cell varies significantly
because of formation of superstructures as displayed in
table~\ref{Table1}. In comparison, the simultaneous changes of the
$a$ and $b$ cell parameters are rather negligible and the crystal
unit cell size change is related to integer multiplications of the
original CeCoAl-type unit along the $c$ axis in agreement with
Ref.~\onlinecite{Feyerherm2012}. Within this process, the number of
inequivalent Ce lattice sites in the intermediate (275~K) and low
temperature (120, 200~K) phase, respectively, is larger than in
the room temperature polymorph (300, 290~K). The structure at
300~K (Fig.~\ref{Fig5}) is practically identical to that one at
290~K and in accordance to previous
reports.\cite{Riecken2007,Feyerherm2012} Upon cooling below the
first transition to $T=275$~K, a tremendous prolongation of the
$c$ axis is observed. The resulting superstructure can be
described as a quintuple of the CeCoAl subcell with five
crystallographically independent cerium sites (see
Fig.~\ref{Fig5}). With further cooling down to 120~K, \CRS\ passes
through the second polymorphic transition at \Tcl\,. The final,
and only existing structure can be viewed as a tripling of the
CeCoAl unit--cell exhibiting three different cerium sites
(Fig.~\ref{Fig5}). The nearest Ce--Ru distances
for all phases are summarized in table~\ref{Table2}. With closer
inspection of the crystallographic parameters, one can see that
all the inter--atomic distances were more or less modified. But
the crucial aspect seems to be the increased number of Ce
positions with short Ce--Ru pairs (cf. table~\ref{Table2}). This
explains the $c$ axis contraction, since Ce--Ru bonds are oriented
almost entirely along this direction and scales the distances
along the $c$ axis. Calculating the relative change of the average
CeCoAl subcell from the obtained lattice parameters yields a
shrinking of about 0.47~\% (between room temperature and 275~K)
and 0.70~\% (between room temperature
and 120~K) along the $c$ axis, which is in line with the thermal expansion results.\\

%===discussion

\section{Discussion}

\begin{figure}[ht]
\centerline{\includegraphics[width=0.8\textwidth]{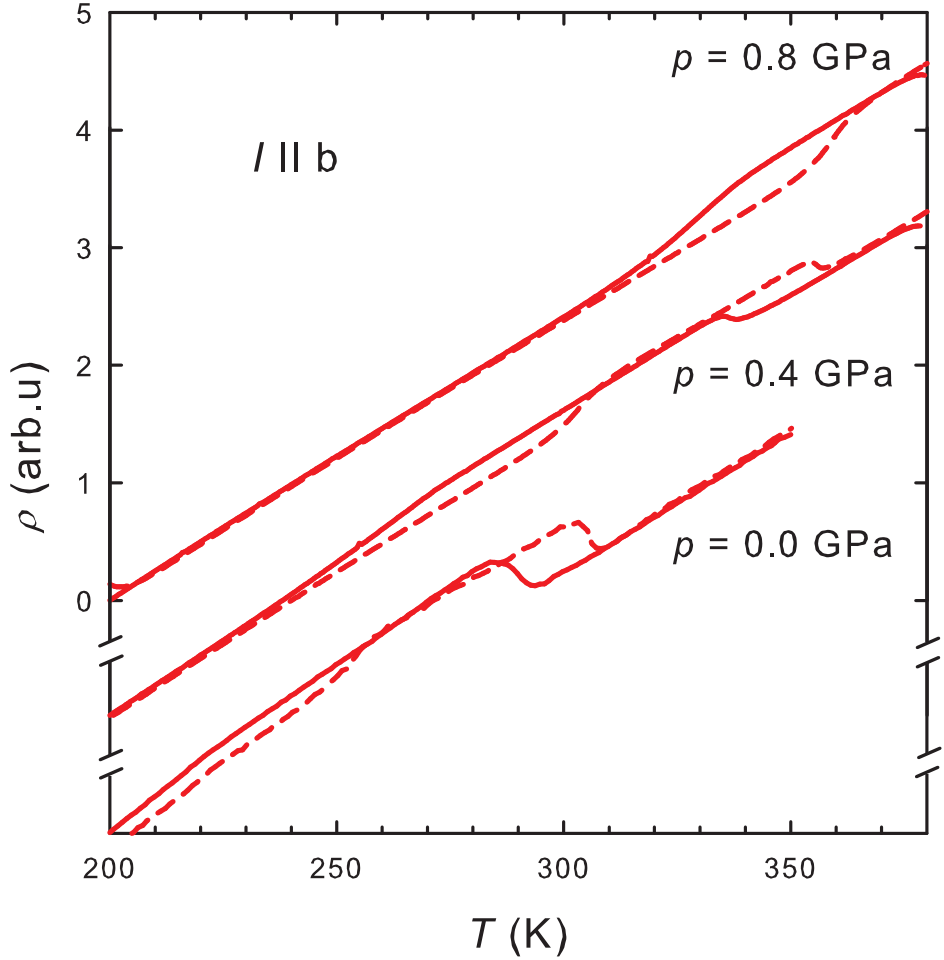}}
%\centerline{\includegraphics[width=\columnwidth]{FikacekCeRuSnFigure6.eps}}
\caption{\label{Fig6} Temperature dependency of the resistivity of
CeRuSn at ambient and under hydrostatic pressure $p = 0.4$~GPa and
0.8~GPa. The current is applied along the $b$ axis. Solid
(short--dashed) lines show measurements in cooling down (warming
up). For $p = 0.8$~GPa, the upper transition is shifted beyond the
maximum temperature of our experiment already.}
\end{figure}

% figures that $B_{\mathrm{infl}}$
\begin{figure}[ht]
\centerline{\includegraphics[width=0.8\textwidth]{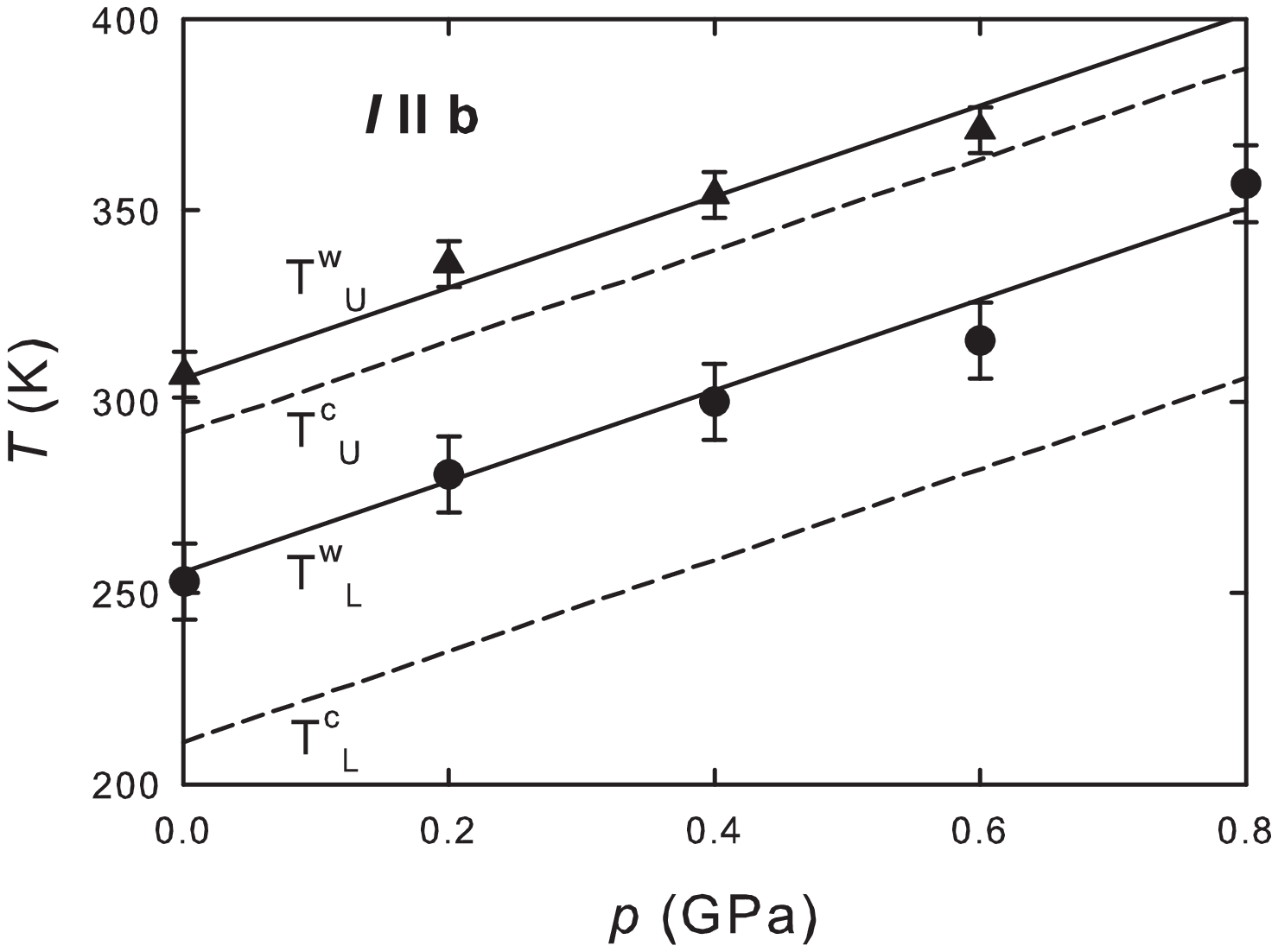}}
%\centerline{\includegraphics[width=\columnwidth]{FikacekCeRuSnFigure7.eps}}
%\centerline{\includegraphics[width=0.8\textwidth,angle=-90]{fig1.eps}}
\caption{\label{rho} $p$--$T$ phase diagram of CeRuSn determined
by resistivity experiments under hydrostatic pressure. The full
lines are the guides for the eye separating the high-temperature,
intermediate and low-temperature polymorphic phase, respectively,
for the warming regime. The dashed lines represent the alternative
for the cooling regime.}
\end{figure}

Signatures of the polymorphic transitions were presented for
resistivity, magnetization and thermal expansion. We did not show the
data on the thermoelectric power, temperature evolution of the
Hall effect and thermal conductivity in which also a step-like
structure was observed at the at \Tcu\,, \Tcl\,, \Twl\, and \Twu\,.
All these findings strongly indicate that the polymorphic
transition involves a Fermi surface reconstruction accompanied by
a change of the electronic structure. A strong electron--lattice
coupling is to be expected. Moreover, as pointed out in the
introduction, because of mutual interplay between electronic
structure and interatomic forces, various lattice vibration
properties can be expected according to whether the 4{\it f\/}
moment is localized in an intermediate state. It was discussed by
Mydosh {\it et. al.}~\cite{Mydosh2011} that the step--like
reduction of the magnetization signals a decrease of the density
of states, i.\,e.\,, change of the electronic structure. However,
one might interpret the observed reduction in $\chi(T)$ (see
Fig.~\ref{Fig2}) as manifestation of a sudden (partial) Kondo
screening of the localized Ce2 moments in \CRS\. The slightly
higher than expected resistivity values for $I \parallel b$ below
\Tcl\, therefore would be a result of a reduction of conduction
electrons involved in screening. Such scenario applies to Cerium
where lattice vibrations are suggested to play an important role
in the Ce $\gamma \rightarrow \alpha$ transition together with
spin and charge degrees of freedom.\cite{Jeong2004} \\
Speculation about an analogy with elementary Cerium are further
inspired by hydrostatic pressure experiments on \CRS\. The upper
accessible temperature (380~K) is limited by the properties of the
StyCast epoxy used for sealing of the wires in the plug of the
pressure cell. In Fig.~\ref{Fig6}, the resistivity data in
arbitrary units are depicted against temperature at ambient and at
pressures of $p = 0.4$ and $0.8$~GPa. The character of the
transitions remains qualitatively the same even at highest applied
pressure when they are still observable within the temperature
range of the experiment. Interestingly, within the applied
pressure range and the resolution of the experiment \Tcu\,,
\Tcl\,, \Twl\, and \Twu\ increase roughly linear with an identical
rate. The slopes of the respective polymorphic transition shift
amounts to approximately 125~K/GPa. This linear increase of the
transition temperature is reminiscent to the linear shifting of
the $\gamma \rightarrow \alpha$ phase line of
Cerium.\cite{Gschneider1962} Here the slope is roughly
250~K/GPa~\cite{Koskenmake1978} twice the ones in \CRS\,.

The results of the structural investigation can be understood as a
subsequent annealing and consecutive evolution of polymorph
phases. In the determined phases, the cerium position (see
table~\ref{Table2}) can be divided into three groups -- those
exhibiting two short $(S)$ Ce--Ru distances (close Ru nearest
neighbors), those with two long $(L)$ Ce--Ru distances (far
nearest neighbors) and those $(I)$ with one close and one far Ru
nearest neighbor (within each group, there is a variation of the
Ce--Ru distance across the polymorphs but in comparison to the
short-long distance the change is minimal). With this in mind (see
Fig.~\ref{Fig5}), the low temperature structure can be described
as $SLLS$ base building block (two CeCoAl subcells, same sequence
as at room temperature) alternating with the $II$ cerium sequence
(leading to the tripling of the CeCoAl subcell observed at low
temperatures) along the $c$ axis. With increasing temperature,
a rearrangement by displacive transformation leads to the extinction
of half of the $II$ spacers leaving the structure with quintuple
CeCoAl subcell (two base blocks, one spacer). Further heating up
removes rest of the $II$ spacers and resulting in the appearance
of the base building block at high temperatures. Within this
context, the existence of the $II$ spacer can be understood as a
deformation of the CeCoAl--sized cell driven by cohesion forces in
order to stabilize the whole structure. The results of the XANES
experiment~\cite{Feyerherm2012} seem to be in contradiction with
the above presented structural data (with decreasing temperature
the number of sites with short Ce--Ru distance is increased).
However, it is necessary to keep in mind, that the dependence of
the valence on the Ce--Ru distance is not simple and that there
are several different short Ce--Ru distances at lower
temperatures. This nonlinearity together with an increased number
of crystallographically inequivalent of Ce sites and presumable Kondo
screening leads to shrinking of the $c$ axis concurrently and to
an unchanged overall Ce valence, which is different from the
statement~\cite{Feyerherm2012} that the valence of the Ce ions
through the transition remains conserved. Further experiments
resolving this issue are desired.\\

To find the true nature, i.\,e.\,, the driving mechanism behind
the polymorphic transitions is a challenging task for future work.
The transitions in \CRS\ show to some extend similarities to the
cerium case, which might serve as reference point. In order to
enlighten the role of lattice vibrations, inelastic neutron
scattering experiments are envisaged.

\section{Summary}
Investigation of the polymorphic transitions by means of
resistivity, magnetization, thermal expansion and X--ray
diffraction on single crystals of \CRS\ was carried out.
Measurements were conducted along all principle axes. In all
physical properties, upon cooling, two subsequent anomalies at
\Tcu\ $\sim 285$~K and \Tcl\ $\sim 185$~K were detected. These
signatures can be attributed to polymorphic transitions, i.\,e.\,,
from the room temperature double CeCoAl-type superstructure to a
quintuple at \Tcu\ and from the quintuple to a triple CeCoAl
unit-cell superstructure at the lower transition temperature. The
refined superstructures are characterized by an increased number
of crystallographically inequivalent Ce sites. Simultaneously, the
ratio between the number of short and long Ce--Ru bonds, which are
essentially aligned along $c$ direction, is increased. As
consequence, the lattice gradually contracts mainly along the $c$
axis as observed in thermal expansion eliciting an overall
shrinking of the sample volume. The transitions exhibit
large hysteric behavior. \\
The strong response of the polymorphic transitions in transport
and magnetic properties infers a close connection to variations in
the electronic structure of \CRS\. Slight jumps in the
magnetization as well as unexpected behavior in resistivity
suggest influence of Kondo interaction to play a role in the
structural change. Together with lattice vibrations, it might be
the driving mechanism behind the polymorphic transitions similar to the one in
elementary Cerium. This scenario is partially rooted in
resistivity data on \CRS\ under hydrostatic pressure revealing an
almost linear increase of the transition temperatures upon
pressure as had been observed for the $\gamma \rightarrow \alpha$
transition in Ce as well.

\section{Acknowledgments}
This work was supported by the Czech Science Foundation (Project
\# 202/09/1027) and Charles University grants GAUK440811 and UNCE
11.

%We gratefully acknowledge discussions with M.\ M,  N.\ N, and A.\
%A. This work was supported by the Grant n$^{\circ}$~227378 (MM)
%and by NSF DMR 0907 (AA).

\vspace{0.5cm}

\noindent${\ast}$ corresponding author: fikacekjan@seznam.cz

\bibliographystyle{prsty}

\end{document}